\documentclass[prd,preprint,superscriptaddress,showpacs,byrevtex]{revtex4}

\usepackage{epsfig}

\newcommand{\be}{\begin{equation}}
\newcommand{\ee}{\end{equation}}
\newcommand{\bea}{\begin{eqnarray}}
\newcommand{\eea}{\end{eqnarray}}
\def\simgt{\rlap{\lower 3.5 pt \hbox{$\mathchar \sim$}} \raise 1pt
  \hbox {$>$}}

\begin{document}
\title{Higgs Production and Decay from TeV Scale Black Holes at the LHC }

\author{Arif Emre Erkoca} \email{aeerkoca@physics.arizona.edu}
\affiliation{Department of Physics, University of Arizona,
Tucson, AZ 85721, USA}
\author{Gouranga C. Nayak} \email{nayak@physics.arizona.edu}
\affiliation{Department of Physics, University of Arizona,
Tucson, AZ 85721, USA}
\author{Ina Sarcevic} \email{ina@physics.arizona.edu}
\affiliation{Department of Physics, University of Arizona,
Tucson, AZ 85721, USA}
\affiliation{Department of Astronomy and Steward Observatory, 
University of Arizona, Tucson, AZ 85721, USA }

\date{\today}
\begin{abstract}

We perform detailed study of the Higgs production and decay, when Higgs is emitted from
 the black holes produced in proton-proton collisions at the Large Hadron Collider.
We show that black hole production can significantly enhance the signal for the
Higgs search at the LHC.
We evaluate 
rapidity distribution of diphotons and 
transverse momentum distribution of 
 bottom quarks, photons, tau leptons, top quarks and 
W bosons from 
 Higgs decay, when Higgs is emitted from the
 black hole and also in case when these particles are produced directly from the
black hole evaporation.
 We compare our results with the standard model backgrounds.  
We find that Higgs production from black holes is
dominant over standard model production for $p_T^H > 100$ GeV, when $M_P=1$TeV.  
Diphotons from Higgs, when Higgs is produced from evaporation of 
black holes, are dominant over the
standard model prediction, for diphoton rapidity $|y_{\gamma \gamma}| 
\leq 1$, while bottom quarks 
are dominant over QCD background 
 for large bottom quark transverse momentum, 
$p_T^b > 300$ GeV, when $M_P=1$ TeV.  
We show that measurements of the
photon and bottom quark transverse momentum distribution can provide
valuable information about the value of the fundamental Planck scale.  
We also propose a new signal for black hole production at the LHC, an onset of 
increasing transverse momentum distribution of bottom quarks with large transverse 
momentum.  

\end{abstract}

\maketitle
\newpage
\section{Introduction}

An idea for solving the so-called hierarchy problem was
proposed by assuming the existence of large
extra dimensions, in which only the gravity can propagate \cite{Arkani}.  
In this model, all ordinary matter is restricted to reside on a 3+1-dimensional brane and
the only fundamental energy scale is taken to be of the order of 
TeV.  In the presence of these large extra dimensions, the gravitational
interaction between
two massive particles can be modified through the 
definition of a higher dimensional Planck scale 
($M_P$), which is related to the four dimensional Planck scale ($M_{Pl}$) by 

\begin{equation}
M^2_{Pl} = {M^{n+2}_{P}}(2\pi R)^n
\end{equation}

\noindent 
with $R$ being the size of each of these extra dimensions 
and $n$ being the number of extra dimensions. When $M_{P}=1TeV$ and 
$n$ is ranging from 2 to
7, $R$ extends from $1$ mm to $1$ fm \cite{Landsberg2}.  
The 
collider experiments and astronomical data have 
put constraints on the parameters 
$n$ and $M_{P}$ \cite{Anchordoqui,Hagiwara}.  For example, supernova cooling 
and neutron star heating by decay of gravitationally trapped 
Kaluza-Klein modes provide limits on $M_P$: $M_P \gg$ 1500 TeV for $n =2$, 
 $M_P \gg$ 100 TeV for $n =3$, 
 whereas for $n=4$ ($n=5$) supernova cooling yields $M_P>4$ TeV 
($M_P>0.8$ TeV). In addition, nonobservation of black hole production by cosmic neutrinos 
provides the most stringent limit on $M_P$, 
i.e $M_{P}>{1-1.4 TeV}$ for $n\ge5$.  

One of the most striking consequences of a low fundamental Planck scale is the 
possibilty of producing 
black
holes and observing them in future 
 colliders or in cosmic rays/neutrino interactions 
\cite{Argyres,Giddings,Cavaglia,Emparan,Myers,Dimopoulos,Han,Meade}.  
At CERN's Large Hadron Collider (LHC) and in high energy 
cosmic ray and neutrino interactions, the collision of particles with 
trans-Planckian energies, i.e. energies larger than $M_P$, is achieved.  
The impact parameter of subatomic particles in this collision 
 can be smaller than the n-dimensional Schwarzschild radius, thus 
making it possible for mini black holes to be produced.  
The black hole production cross section can be approximated with the 
geometrical cross section, 
$\sigma\sim\pi{b}^2\sim\pi M_P^{-2}$, which corresponds to 
 about $100{\rm pb}$ for $M_P=1$TeV.   
When the luminosity at the LHC reaches its peak value of 
$L=10^{34}cm^{-2}s^{-1}$, this cross section corresponds 
to several million of events with black holes in one year of running.

Semi-classical approach restricts us to explore the parameter space 
where the mass of the black hole, $M_{\rm BH}$, is much larger than 
 the fundamental Planck mass,
$M_P$.  Around and below this mass scale, quantum gravity 
effects become important.  We will explore only the parameter space 
where the semi-classical approach is justified.  

Once the black hole with mass larger than $M_P$ is formed it quickly
evaporates via Hawking radiation \cite{Hawking}.  Since these black holes are very small in
size ($\sim$ a few $10^{-4} fm$ ), their temperatures are 
very high ($\sim $ 500 GeV - 1 TeV).  Thus, they can emit 
heavy particles with masses less than their 
temperatures. The emission spectrum of these particles depends 
not only on their spin properties and energies but also on the structure of the spacetime.  
These effects can be desribed by the so-called 
``greybody factors'' which define the effective area of the emitting body.
 Greybody factors were obtained for 
scalars, fermions and gauge bosons in the case for non-rotating \cite{Kanti}
and rotating black holes \cite{Creek}. 

One of the primary goals of the LHC is to search
for the Higgs, particle which is responsible for generating masses of the 
particles in the Standard Model.  
 The dominant channel for 
Higgs production in p-p collisions  
at LHC energies is via gluon fusion process.   Subdominant contributions come from 
the quark antiquark and vector boson scatterings \cite{Spira}.
Since Higgs can not be
detected directly at LHC
the primary decay channel proposed for Higgs detection with ATLAS and 
CMS detectors is via diphoton production \cite{CMS}.  
A discovery significance above 5 sigma is expected at LHC for Higgs with mass
below 140 $\rm GeV$ for an integrated luminosity of 30 $\rm fb^{-1}$.  
Higgs is expected to have mass above $114.4$ $\rm GeV$, which is the lower limit from 
 the LEP data \cite{Barate}.  The Higgs discovery 
signal which consists of two isolated
photons with high transverse momenta can be identified as a 
narrow peak at the Higgs mass on top of a continuous  background.
Other Higgs decay channels such as bottom quarks and tau leptons 
 may be measured at LHC as well \cite{Loch}. Compared to the diphoton channel, 
these decay channels have larger decay branching fractions in the range of
Higgs mass considered, but they are experimentally
 challenging due to the presence of large QCD background \cite{Ganjour}.

Recently a new mode of Higgs production via Hawking radiation of 
black holes at LHC has been proposed \cite{Dimopoulos}.  In this paper 
 we make a detailed calculation of the various decay 
channels of the Higgs emitted from the black hole, 
such as diphotons, bottom quarks, tau leptons, W bosons and top
quarks.  We also present results for the production of the same 
final state particles produced from the black hole evaporation.  
We compare our results with the 
corresponding Standard Model background processes.  

The paper is organized as follows. In section II we describe production and the
decay of black holes at LHC, including Higgs production from black holes.
In section III we describe the decay of the Higgs which is produced from black holes.
In section IV we present the results and discussions and we conclude in section V.

\section{Production and Decay of Black Holes at LHC}

In the theories of large extra dimensions gravitational effects become stronger 
at distances which are less than the size of
the extra dimensions and at 
trans-Planckian energies,
mini black hole production becomes possible.  
The radius of the black holes depends on the mass of the black hole, 
number of extra dimensions and the fundamental Planck scale, 
\cite{Myers}

\begin{equation}
R_{\rm BH}=\frac{1}{\sqrt{\pi}M_P}\left(\frac{M_{\rm BH}}{M_P}
\frac{8\Gamma((n+3)/2)}{n+2}\right)^{\frac{1}{n+1}}, 
\end{equation}

\noindent 
The temperature of the black hole is inversely proportional to its radius, i.e. 

\begin{equation}
T_{\rm BH}=\frac{n+1}{4\pi{R_{\rm BH}}},
\end{equation}

\noindent 
and its decay time is given by 

\begin{equation}
\tau_{\rm BH}=\frac{1}{M_P}\left(\frac{M_{\rm BH}}{M_P}\right)^{\frac{n+3}{n+1}}.
\end{equation}

Black hole decays emitting elementary particles with masses smaller than the
temperature of the black hole, $T_{BH}$, via Hawking radiation,
a process which has never been observed for astronomical black holes.
The number of particles emitted from the black hole per unit time
is given by \cite{Kanti}:

\begin{equation}
\frac{dN}{dt}=\frac{\sigma_G}{exp(E/T_{BH})\mp 1}\frac{d^3p}{(2\pi)^3}
\end{equation}

\noindent
where 
$E$ and $p$ are the energy and the momentum of the emitted particle and 
the spin statistics factor is -1 for bosons
and +1 for fermions, ${\sigma}_G$ is the greybody factor. 
In this paper, we use the geometrical optics limit for  ${\sigma}_G$, i.e \cite{Harris,sanchez,Misner}:

\begin{equation}
\sigma_{G}=\pi\left(\frac{n+3}{2}\right)^{2/(n+1)}\frac{n+3}{n+1}R^2_{\rm BH} .
\end{equation}

The differential cross section for the standard model
particles produced via the decay of the black holes is given by \cite{ina}

\begin{equation}\label{bhdecay}
 E\frac{d\sigma}{d^3{p}}=\frac{1}{(2\pi)^3s}\sum_{a,b}\int^s_{(M^{min}_{\rm BH})^2} d{{M^2_{\rm BH}}}\int^1_{\frac{M^2_{\rm BH}}{s}}
\frac{d{x_1}}{x_1}{f_a}({x_1},Q^2)\hat\sigma{f_b}({x_2},Q^2)
\frac{g{p}^{\mu}{u}_{\mu}{\sigma_G}\gamma{\tau}_{\rm BH}}{exp({p}^{\mu}{u}_{\mu}/{T}_{\rm BH})\mp 1}, 
\end{equation}

\noindent 
where $\hat\sigma$ is the parton level black hole production cross section which can be 
approximated with the pure geometrical form as $\hat\sigma\sim\pi{R^2_{\rm BH}}$,  
$Q$ is the factorization scale, $f_a$ and $f_ b$'s are the parton
distribution functions, $x_1$ and $x_2$ are the momentum fractions of the 
colliding partons which satisfy $x_1x_2s=M^2_{BH}$, $\sqrt{s}$ is the center of mass energy of the
colliding
hadrons, $M^{min}_{\rm BH}$ is the minimum black hole mass, 
 $\gamma$ is
the Lorentz factor, $g$ is the number of internal degrees of freedom,
and $p^\mu$ is the four momentum of the emitted particle 

\begin{eqnarray}
p^{\mu} & = & (\sqrt{m^2+p^2_T}cosh(y),p_x,p_y,\sqrt{m^2+p^2_T}sinh(y))  
\end{eqnarray}

\noindent
and $u^\mu$ is the 
four velocity of the black hole, 
${u}^{\mu}  =  (\gamma,0,0,\gamma{v}_{\rm BH})$.  
  Clearly ${u}^{\mu}$ satisfies ${u}^{\mu}{u}_{\mu}=1$.  
  In the center of mass frame, 
$\vec{p}_1+\vec{p}_2=0$
and the momentum of the black hole becomes

\begin{equation}
\vec{P}_{\rm BH}=(x_1-x_2)\vec{p}_1 ,
\end{equation}
where $\vec{P}_{\rm BH}={\gamma}{M_{\rm BH}}\vec{v}_{\rm BH}$ and $\vec{p}_1$ and $\vec{p}_2$ are
the momenta of the colliding hadrons.  
If we assume that ${p_1}^{\mu}{p_1}_{\mu}=0$, and use the fact that 
$\left|\vec{p}_1\right|=\frac{\sqrt{S}}{2}$ we get 

\begin{equation}
 \gamma{v_{\rm BH}}=\frac{(x_1-x_2)\sqrt{s}}{2M_{\rm BH}} .
\end{equation}

We find that the assumption of stationary black hole is 
reasonable because the cross sections 
calculated for the case of the stationary black hole ($v_{BH}=0$) differs only by 
few percent from the cross section obtained by 
incorporating the effect of translational motion along the initial beam axis. 
This is due to the fact that not many black holes can reach
relativistic speeds after they are formed. In addition, we make use of the
identity 

\begin{equation}\label{identity}
E\frac{d\sigma}{d^3p}\equiv\frac{d\sigma}{2\pi{p_T}dp_Tdy}   
\end{equation}

\noindent
to obtain the rapidity distribution,
$\frac{d\sigma}{dy}$
 and the transverse momentum distribution,
$\frac{d\sigma}{dp_T}$,
 by
integrating Eq. (7) for the particles directly emitted from the black hole.

\section{Decay of Higgs produced from evaporation of Black Holes}

To evaluate differential 
 distributions 
of standard model particles from Higgs decay when the 
 Higgs is emitted from the black hole, we 
 fold the decay distribution of the Higgs with the
differential decay rate 
for the decay mode considered: 

\begin{equation}\label{diffx1}
E_1E_2\frac{d\sigma}{d^3p_1d^3p_2}=\int\frac{d^3p}{E}
\left(E\frac{d\sigma}{d^3p}\right)_{\rm BH\rightarrow{H}}
\left(\frac{B}{\Gamma}E_1E_2\frac{d\Gamma}{d^3p_1d^3p_2}\right)_{\rm H\rightarrow{1+2}}, 
\end{equation}

\noindent
where $\Gamma$ is the decay rate, $B$ is the decay branching fraction, 
subscripts 1 and 2 refer to the decay products, $E$ and $p=|\vec{p}|$ are the energy and the momentum
of the Higgs, 
respectively ($E=\sqrt{p^2+m^2_H}$).  From Eq. (\ref{diffx1}), one can obtain the 
differential cross section for one of the decay products with mass m:

\begin{equation}\label{diffx2}
E_1\frac{d\sigma}{d^3p_1} = 
\frac{B}{p_1\sqrt{1-4m^2/m^2_H}}\int^{E_{max}}_{E_{min}}
\left(E\frac{d\sigma}{d^3p}\right)_{\rm BH\rightarrow{H}}dE.  
\end{equation}

\noindent
where $\left(E\frac{d\sigma}{d^3p}\right)_{\rm BH\rightarrow{H}}$ is given by 
Eq. (\ref{bhdecay}). The derivation of Eq. (\ref{diffx2}) is given in the Appendix in which 
we also show that the normalized decay 
distributions agree with those given in \cite{Gaisser}.
This suggests
that regardless of the type of the decay mode, the final expressions (Eq. (\ref{diffx2}))
for the single particle differential cross section always take
the same form.

Furthermore, by considering the change of variables: 

\begin{equation}
Q^{\mu}=p^\mu_1+p^\mu_2\;\;\; \mbox{and}\;\;\; 2k^\mu=p^\mu_1-p^\mu_2  ,
\end{equation}

\noindent 
it becomes possible to solve the integrals analytically in 
Eq. (\ref{diffx1}) which then leads us to a simple 
relation between the differential cross section for the
pair produced from the decay of Higgs emitted from the 
black hole and that for the Higgs produced by the black hole:

\begin{equation}\label{pair}
Q^{0}\frac{d\sigma}{d^3Q}=B\left(E\frac{d\sigma}{d^3p}\right)_{\rm BH\rightarrow{H}}  .
\end{equation}

\noindent
Using Eq. (\ref{identity}) we calculate distributions for the single particles and for the particle pairs
by integrating Eq. (\ref{diffx2}) and Eq. (\ref{pair}), respectively.

\section{Results and Discussions}

In this section we present results of Higgs production
and decay from black holes at LHC. We also show our results for several standard 
model particles produced directly from black hole evaporation. 
Given current experimetal limits, we consider the scenarios with
$n=6$ and $M_P=1 - 3$ TeV. Since quantum effect become important when the black hole mass 
is 
close to the Planck mass, we consider black holes with mass 
 $M_{\rm BH}^{\rm min}~\ge ~3~ M_P$.
We check that black hole production cross section at LHC is not sensitive to the choice for the number
of extra dimensions.  However, we find that our results depend on the value of the
Planck mass and the minimum black hole
mass. We use CTEQ6M parton distribution \cite{cteq} with the factorization
scale equals to ${R_{\rm BH}}^{-1}$ where $R_{\rm BH}$ is the black hole radius.
We compare our results with the corresponding standard model predictions and also discuss the 
dependence of our results on
the parameters $n$ and $M_P$.

\subsection{Higgs Production at LHC}

In Fig. 1 we present the total cross section for Higgs production from black holes
as a function of the Higgs mass at LHC and we compare our result with that of the 
next-to-next-to-leading order pQCD \cite{smith} 
obtained in the Standard Model.
We find that for the Planck mass $M_P=1$TeV, and the
minimum black hole mass,
$M_{\rm BH}^{\rm min}$=3 $M_P$,
the Higgs production cross section from black holes at the LHC is much larger than
the Standard Model prediction. This is qualitatively consistent with the result in
\cite{Nayak}, which was obtained with fixed black hole mass.
Therefore, if black holes are produced at LHC, it will
have a direct impact on Higgs search at LHC.

In Fig. 2 and Fig. 3 we present transverse momentum distributions of the Higgs, when Higgs is 
produced from black holes, compared with 
the Higgs produced via 
standard model processes obtained using next-to-leading order 
 pQCD \cite{smithr}. We note that transverse momentum distribution
of Higgs from black holes increases as $p_T$ increases, in sharp contrast
to the result obtained from the standard model predictions where
the transverse momentum distribution
 decreases
as $p_T$ increases.  As we will exemplify later, this is not only true for Higgs production from black holes
but also for any particle emitted from black holes. 
Therefore, if TeV scale black holes are produced at LHC
then an important signature is the increase of
the transverse momentum distribution
 as $p_T$ increases.  The origin of this $p_T$ dependence is
the high black hole temperature ($\sim$ 500 GeV -1 TeV).
If black holes are indeed produced at LHC, then the black hole physics will
dominate over all standard model physics. 

In addition, our results indicate that the Higgs production
via Hawking's radiation
crucially depends on the value of the Planck mass and the minimum black hole mass and only marginally on the number of 
extra dimensions. As a numerical example,
 increasing $M_P$ from $1$ TeV to $2$ TeV and $M_{\rm BH}^{\rm min}$ from $3$ TeV to $6$ TeV
 results in a decrease of the cross section by two orders of magnitude. A further suppression, no more than $30\%$,
 also occurs when $n$ is decreased from 6 to 5.
 The shape of the distribution, however, remains unchanged.

\subsection{Higgs Decay at LHC}

Since the Higgs can not be directly detected at the LHC, we focus on 
 various decay modes of the Higgs emitted by the black holes such as diphotons, bottom quarks, tau leptons, W bosons and top quarks. We
compare these results with the standard model predictions. We use expressions
for various Higgs decay mode distributions derived in the previous section.

We first concentrate on the diphoton production. We find that the total cross 
section for diphoton 
production from Higgs is increased by a factor of five due to the black hole contribution,
when compared to the Higgs decay into diphoton in the Standard Model \cite{Dixon}. This enhancement also becomes apparent in the 
normalized transverse momentum distribution (Fig. 4) and in the normalized rapidity distribution (Fig. 5). We find that in the central rapidity region,
$-1 \leq y \leq 1$,
 and for $p_T>$ 100 GeV, the contribution from the Higgs emitted from the black hole becomes dominant. Thus,
the formation of TeV scale
black holes at LHC can significantly impact Higgs search via
diphoton measurements with ATLAS and CMS detectors at LHC.
In this study, we also analyze the effect of the black hole on 
the transverse momentum distribution for single photon. We find that
$p^{\gamma}_T>$ 1 TeV region needs to be explored to observe this effect
which dominates over the Standard Model background 
given in \cite{Blair}. 
However, the direct 
photon production from the black hole can be dominant over the Standard Model processes even for $p_T>$ 250 GeV (Fig. 6).

We also consider other decay channels of Higgs, such as 
$H\rightarrow{b}{\bar{b}}$ and $H\rightarrow{\tau^+\tau^-}$.  
In Fig. 7 we present 
transverse momentum distribution 
for bottom quarks produced both directly and via Higgs decay from black holes and we compare our results with that of
 next-to-leading order pQCD  \cite{Nason}. We find that 
the bottom quark production is suppressed by a few orders of magnitude when $M_P$ is increased 
from 1 TeV to 2 TeV.  For $p_T>$ 200 GeV bottom quarks emitted from black holes dominate over 
the ones produced from Higgs, when $M_P=1$ TeV and for $M^{min}_{BH}=3$ TeV.  However, the background 
from pQCD processes \cite{Nason} is significant for bottom quarks with transverse momentum 
below $300$ GeV.  When $M_P=2$TeV and 
 $M_{BH}>6$ TeV, the background remains dominant to even larger values of the 
transverse momentum.    
To illustrate dependence of the bottom quark transverse momentum distribution on the choice of 
$M_{BH}^{min}$, 
we show in Fig. 7 the 
 bottom quark
production from black holes via Hawking's radiation for different minimum black hole masses
with Planck mass being fixed to 1 TeV.  
We find that the shape of the bottom quark distribution is the same, but the overall 
normalization is about one order of magnitude smaller for the case of 
$M_{BH}^{min}=5$ TeV than it is for $M_{BH}^{min}=3$ TeV .

In Fig. 8 we summarize our results including the 
$\tau$ lepton transverse momentum distributions for both direct and indirect (via Higgs decay) production from the black hole.

In summary, we note that in general 
there is an onset in the transverse momentum distributions for the particles 
produced directly from the black holes below $500$ GeV of $p_T$ regardless of the choice of $M_P$ or $M^{min}_{BH}$
whereas a sharp decrease in these distributions is expected in the Standard Model scenarios.  
This is due to the high black hole temperatures which greatly exceed 
the mass of the particles considered. This feature can be used as a potential signal for the black hole production at LHC.   
On the other hand, the cross 
sections for the decay products of Higgs emitted from the black holes are slowly decreasing with 
$p_T$ which can still contribute more than the Standard Model predictions.  
Furthermore, for the direct production of the particles from the black hole, we observe that the shape of the distributions mainly depend on the spin 
statistics. In addition, for a given statistical distribution, the overall normalization turns out to be dependent mostly on 
the number of internal degrees of freedom, $g$, whereas the effect of the particle mass seems insignificant especially at high $p_T$. As an example, 
when 
$p_T$=500 GeV, $(\frac{d\sigma}{dp_T})_{\gamma}/(\frac{d\sigma}{dp_T})_{Higgs}=3.139 \sim 3$ which is the ratio of the number of internal degrees of 
freedom for these particles.

\subsection{Parametrization}

Our results from the previous sections seem to present a universal behavior in
the transverse momentum distributions not only for particles produced directly
from the black hole (Set 1) but also for the decay products of Higgs emitted from the
black hole (Set 2). For Set 1, the shapes of the distributions turned out to be mostly dependent on the
spin statistics factor, and the overall normalizations on the number of internal degrees of freedom for a given spin statistic.
From the analysis of Set 2, we observe that the main difference between 
the distributions is due to the distinct branching fractions of the Higgs decays as presented below: 

In order to parameterize our results, we fit each
resulting distribution with
the polynomial function in the form

\begin{equation}
f(p_T)=N\times\left[A_0+A_1\times{p_T}+A_2\times{p^2_T}\right].
\end{equation}
 
\noindent
where $N,A_0,A_1$ and $A_2$ are the best fit parameters.
In order to find them, we apply Chi-Squared statistics together with
LEVENBERG-MARQUARDT minimization method. Table (1) and Table (2) present the values of these
parameters. As it is seen from Table (1), we obtained the parameters for Set 1, at best, by considering two different
momentum ranges (i.e $p_T<500$ GeV and $p_T>500$ GeV). However, we do not need to follow this procedure for Set 2.
As a result, all the fits turn out to possess percentage errors $<5\%$ in particular for $p_T>100$ GeV below which they tend to
become larger due to the effect of the particle mass.

\begin{table}[h]
\centering
\begin{tabular}{lr|ccc|ccc}
\hline\hline
\multicolumn{2}{c}{Fit Parameters}&\multicolumn{3}{c}{bosons}&\multicolumn{3}{c}{fermions} \\[-1ex]
\hline
 & $A_0 (pb/GeV)$ & & $1.823\times10^{-2}$ & & &$-1.560\times10^{-2}$& \\
$p_T<500 GeV$ & $A_1 (pb/GeV^2)$ & & $5.425\times10^{-4}$ & & &$2.903\times10^{-4}$&  \\
 & $A_2 (pb/GeV^3)$ & &$-5.786\times10^{-7}$& & &$-1.819\times10^{-7}$& \\
\hline
 & $A_0 (pb/GeV)$ & &$1.358\times10^{-1}$& & &$-1.540\times10^{-2}$& \\
$p_T>500 GeV$ & $A_1 (pb/GeV^2)$ & &$6.185\times10^{-5}$& & &$2.755\times10^{-4}$&  \\
 & $A_2 (pb/GeV^3)$ & &$-7.318\times10^{-8}$& & &$-1.630\times10^{-7}$& \\
\hline\hline
 & &photon&W&Higgs&tau&bottom&top \\[-1ex]
\hline
\multicolumn{2}{c}{$N(p_T<500 GeV)$} &3.00&2.94&0.96&2.00&6.00&5.95 \\
\multicolumn{2}{c}{$N(p_T>500 GeV)$} &3.00&2.98&0.98&2.00&6.00&5.88 \\
\hline
\multicolumn{2}{c}{Percentage Errors}&1\%&1\%&2\%&2\%&2\%&5\% \\
\hline
\end{tabular}
\caption{Fit parameters for the transverse momentum distributions of the particles produced directly from black holes.}
\end{table}

\begin{table}[h]
\centering
\begin{tabular}{l|ccc}
\hline\hline
 $A_0 (pb/GeV)$   & \multicolumn{3}{c}{$3.899\times10^{-1}$}  \\
 $A_1 (pb/GeV^2)$ & \multicolumn{3}{c}{$-5.648\times10^{-4}$}   \\
 $A_2 (pb/GeV^3)$ & \multicolumn{3}{c}{$2.541\times10^{-7}$}  \\
\hline
&photon&tau&bottom \\[-1ex]
\hline
$N$ &0.00221&0.0629&0.684 \\[1ex]
\hline
Percentage Errors&2\%&2\%&2\% \\ [1ex]
\hline
\end{tabular}
\caption{Fit Parameters for the transverse momentum distributions for the decay products of Higgs emitted from the black hole.}
\end{table}

According to the values of the best fit parameters and the corresponding errors, we can conclude that the distributions indeed follow a general trend
,while, the effect of particle mass becomes important
at relatively low $p_T$ and the fits have smaller percentage errors for particles with low mass.
 We also observe that the normalization parameters for
Set 1 are just the number of internal degrees of freedom of the   
particles emitted from the black hole as noted in the previous section. Furthermore, we also
find that the normalization parameters for Set 2 are
the decay branching fractions of the modes considered in this paper. Our results are obtained for 
the model parameters: $n=6$ , $M_P=1$ TeV and $m_H=120$ GeV and in the 
transverse momentum range, $p_T<1$ TeV.

\subsection{W Boson and Top Quark Production From Heavy Higgs, when Higgs is Emitted from Black Holes }

When the Higgs is heavier than about $150$GeV, its dominant decay channels are 
$W^{\mp}$ (and $Z^0$) bosons and 
top quarks.
In Fig. 9 we present 
transverse momentum distributions for $W$ boson production
from black holes and from the Higgs, when Higgs is emitted from black holes.  
 We also compare our results with the pQCD $W$ boson production 
\cite{Wboson}. We take Higgs mass equal to 170 GeV and
$W$ mass to be 80 GeV.
Similar to our previous results, $W$ boson production from black holes is larger than that produced from the Higgs decay.
 The latter distribution increases up to about 100 
GeV and then starts decreasing.
while, the former one
increases as $p_T$ increases. Both distributions dominate over the Standard Model prediction, especially for $p_T>300$ GeV.   
Similar conclusions can also be derived for the 
top quark production (Fig. 10). Here, we again compare our results with that of the
pQCD calculation \cite{topQCD} and we take Higgs mass to be 500 GeV and top quark mass to be 173 GeV in our calculations.

\section{Conclusions}

In this paper we have studied Higgs production and decay, when Higgs is produced
from black holes at LHC.  
If the fundamental Planck scale is near a TeV, then parton collisions with
high enough center-of-mass energy at LHC produce black holes.  Since the
temperature of these black holes are very high ($\sim $ TeV), we might
expect to see Higgs production from black holes at LHC via Hawking radiation.
We have made a detailed study of the Higgs production and decay from black holes
at LHC. We have calculated transverse momentum distribution for
diphotons, tau lepton, bottom quarks, W boson and top quark produced
from black holes decay and also from Higgs from black holes at the LHC. We have compared
our results with the corresponding standard model predictions. We have found that for Planck mass
$\ge $ 1 TeV and black hole mass $\ge $ 3 TeV, the diphoton and bottom quark production
from Higgs from black holes at LHC is much larger than the standard model predictions.
Hence black hole production can significantly impact direct Higgs search at LHC.
We have found that 
transverse momentum distribution 
of particle production from black holes
increases as $p_T$ increases.  Therefore we suggest that
a measurement of increase in 
transverse momentum distribution 
as $p_T$ increases is a potential
signature of black hole production at the LHC. 
In addition, we have also noted that the shape of the 
transverse momentum distributions differs depending whether the emitted particle is a boson or a fermion. 
We have discovered that the major effect comes from the number of internal degrees of freedom in determining 
the overall normalizations of the distributions for the 
particles which obey the same spin statistics; however, the particle mass has almost no effect. We have also presented 
some parametric results that can be useful for some other studies.      

In our calculations, we have considered the evaporation of the non-rotating black holes. 
It was shown in \cite{Creek} that the geometrical optics limit for the greybody factors in the case for 
fast rotating black holes is a factor of 3 larger than the one we have used. 
It can also be shown that the temperature of the black holes remains almost the same in the case of rotation. 
Therefore, the inclusion of rotation would lead to an enhancement in our results at most by a factor of 3.        

We have also studied different decay channels of the Higgs which are produced
from black holes at LHC for different values of the Planck mass. A direct
comparison of the differential distributions with the experimental data at LHC will provide 
valuable information about the value of the fundamental Planck mass. Since the diphoton production from Higgs from black holes at LHC
is larger than the standard model predictions, when the fundamental Planck mass is below 3 TeV, it enhances the
chance of detecting Higgs at LHC via diphoton measurements at ATLAS
and CMS detectors.

\acknowledgments We thank Nick Kidonakis and Ramona Vogt for providing us the data for the 
cross section of the top quark in pQCD.
We also thank Tolga G{\"u}ver, Peter Loch and Jack Smith for many useful discussions.
This work was supported in part by Department of Energy under contracts DE-FG02-04ER41319 and DE-FG02-04ER41298.

\appendix
\section{Higgs Decay Rates}

In this section, we present our derivation of Eq. (\ref{diffx2}). We first start with the
 differential decay rate 
distributions of Higgs decaying into any two-body final states
in the laboratory frame. 

Decay distribution for any two-body decay of Higgs, i.e   
$H\rightarrow{X}{\bar{X}}$ is given by 

\begin{equation}
\frac{d\Gamma}{d^3p_1d^3p_2}=\frac{1}{(2\pi)^2}\frac{1}{8EE_1E_2}\delta^3(\vec{p_1}+\vec{p_2}-\vec{p})\delta(E_1+E_2-E)|M|^2
\end{equation}

\noindent 
where 1 and 2 refer to the decay products both of which have mass $m$, 
$|M|^2$ is the spin averaged invariant decay amplitude squared and 
$E$, 
$p=|\vec{p}|$ and $m_H$ are the energy, momentum and the mass of the 
Higgs respectively.  
  After integrating over the momentum $p_2$ we get 

\begin{equation}
\frac{d\Gamma}{d^3p_1}=\frac{1}{(2\pi)^2}\frac{1}{8EE_1E_2}\delta(E_1+E_2-E)|M|^2 \;\;  \mbox{and}\;\;  \vec{p}=\vec{p_1}+\vec{p_2}
\end{equation}

\noindent 
We take the z-axis to be along the direction 
of $\vec{p}$ and write $d^3p_1=2\pi{\sqrt{E^2_1-m^2}}E_1dE_1d(cos\theta)$.  
Using the properties of the delta
function, we solve the angular part of the integration and obtain

\begin{equation}\label{dgdE_f}
\frac{d\Gamma}{dE_1}=\frac{|M|^2}{16\pi{Ep}} \;\; \mbox{and}\;\; E_{-}<\;E_1\;<E_{+}
\end{equation}

\noindent 
where 

\begin{equation}\label{E1range}
E_\mp=\frac{E\mp{p}\sqrt{1-4m^2/m^2_H}}{2}
\end{equation}

\noindent 
Eq. (A3) is a flat differential decay rate distribution with 
respect to the energy of one of the final state particles.  Thus, the momentum and the energy conserving
delta functions restrict the energy range of the final state 
particle for a given energy of the Higgs.

Since $|M|^2$ is independent of $E_1$, it is simple to integrate over the energy $E_1$ and obtain the 
decay rate for a given mode:

\begin{equation}\label{g_f}
\Gamma(E)=\frac{|M|^2}{16\pi{E}p}\Delta{E_1},  
\end{equation}

\noindent
where $\Delta{E_1}$ is the range of $E_1$ that can be calculated by using Eq. (\ref{E1range}).

From the Eqs. (\ref{dgdE_f}) and (\ref{g_f}) we obtain the
normalized decay rate
energy distribution

\begin{equation}\label{dndE_f}
\frac{B}{\Gamma}\frac{d\Gamma}{dE_1}=\frac{B}{p\sqrt{1-4m^2_f/m^2_H}}
\end{equation}

\noindent 
where B is the decay branching fraction of the mode.  
This result satisfies the normalization condition given in \cite{Gaisser}.  
This suggests us that we use  
the decay rate distribution in the following normalized form;

\begin{equation}\label{norm_diff_f}
\frac{B}{\Gamma}\frac{d\Gamma}{d^3p_1}=\frac{B}{2\pi{E_1E_2}}\frac{\delta(E_1+E_2-E)}{\sqrt{1-4m^2/m^2_H}}
\end{equation}

In this paper, we calculate the differential cross section for the 
Standard Model particles produced from the decay of Higgs which is emitted from the 
black holes at the LHC.  
In order to find the single particle distribution we start with the equation

\begin{equation}
E_1\frac{d\sigma}{d^3p_1}=
\int\frac{d^3p}{E}\left(E\frac{d\sigma}{d^3p}\right)_{\rm BH\rightarrow{H}}\left(\frac{B}{\Gamma}E_1\frac{d\Gamma}{d^3p_1}\right)_{\rm
H\rightarrow{1+2}}.  
\end{equation}

\noindent 
With a help of Eq. (\ref{norm_diff_f}) we get 

\begin{equation}
E_1\frac{d\sigma}{d^3p_1}=
\int\frac{d^3p}{E}\left(E\frac{d\sigma}
{d^3p}\right)_{\rm BH\rightarrow{H}}\frac{B}{2\pi{E_2}}\frac{\delta(E_1+E_2-E)}{\sqrt{1-4m^2/m^2_H}}.  
\end{equation}

\noindent 
We assume that black holes produced in the proton-proton collisions 
remain stationary, and the spectrum of emitted particle then is 
 spherically symmetric. 
We 
 define the z-axis along the direction of $\vec{p}_1$ and 
write $d^3p=2\pi{p}EdEd(cos\theta)$ with $\theta$ being 
the angle between $\vec{p}$ and $\vec{p}_1$.  
Using the energy conserving delta function the angular integration leads to

\begin{equation}
E_1\frac{d\sigma}{d^3p_1}=\frac{B}{p_1\sqrt{1-4m^2/m^2_H}}\int^{E_{max}}_{E_{min}}\left(E\frac{d\sigma}{d^3p}\right)_{\rm BH\rightarrow{H}}dE
\end{equation}

\noindent 
where the limits of integration are 

\begin{equation}
E_{min}=\left(\frac{E_1-p_1\sqrt{1-4m^2/m_H^2}}{2}\right)(\frac{m_H}{m})^2
\end{equation}

\begin{equation}
E_{max}=\left(\frac{E_1+p_1\sqrt{1-4m^2/m_H^2}}{2}\right)(\frac{m_H}{m})^2  
\end{equation}

\noindent
for
 $m\neq 0$ and

\begin{equation}
E_{min}=E_1+\frac{4m^2_H}{E_1}
\end{equation}

\begin{equation}
E_{max}=\infty
\end{equation}

\noindent
for
 $m=0$.

\newpage
\begin{figure}[ht]
\begin{center}
\epsfig{file=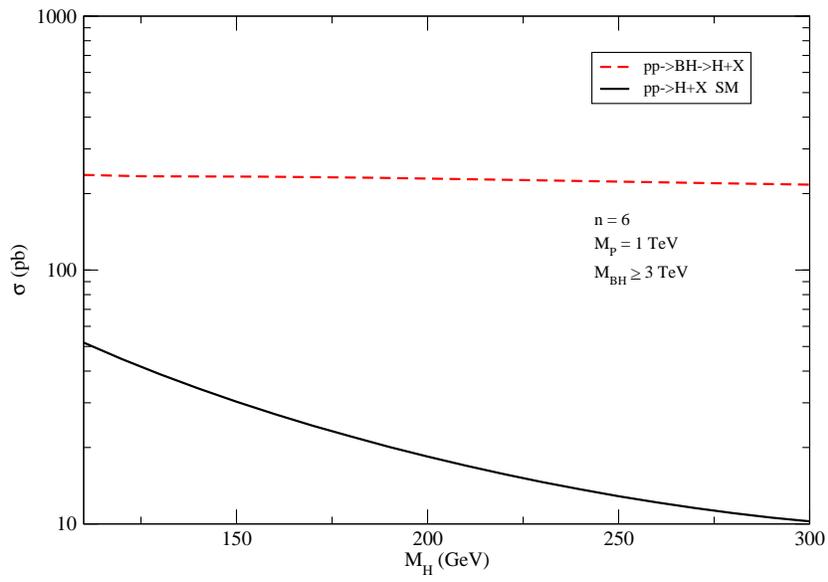,width=3.8in,angle=270}
\end{center}
\caption{Total cross section for the Higgs production from the black holes at LHC (dashed) compared with the standard model prediction (solid) \cite{smith} as a function of Higgs mass.}
\label{}
\end{figure}

\begin{figure}[ht]
\begin{center}
\epsfig{file=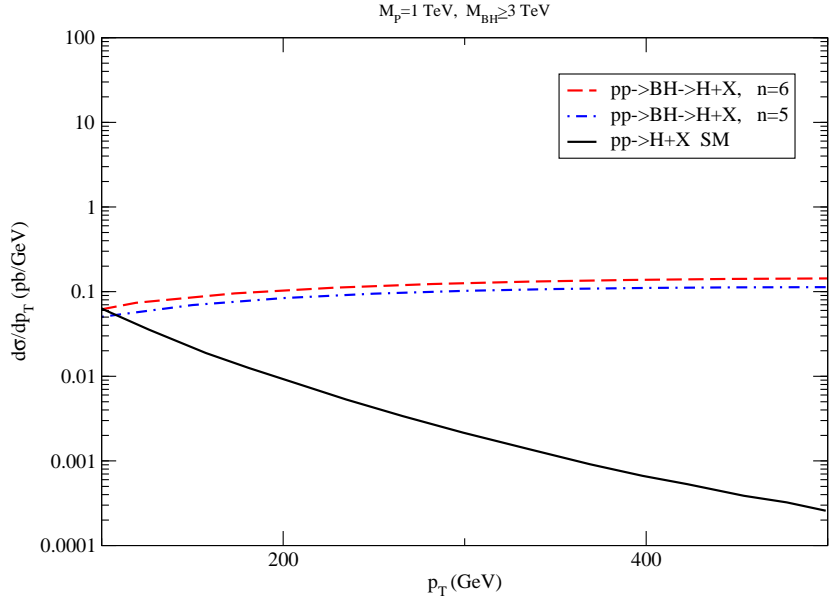,width=3.8in,angle=270}
\end{center}
\caption{Transverse momentum distribution for the Higgs with mass of $120 GeV$ produced by the 
black holes at LHC (dashed line) in a rapidity range $-2.5<y<2.5$ compared with the
standard model result (solid line) \cite{smithr}.}
\label{}
\end{figure}

\begin{figure}[ht]
\begin{center}
\epsfig{file=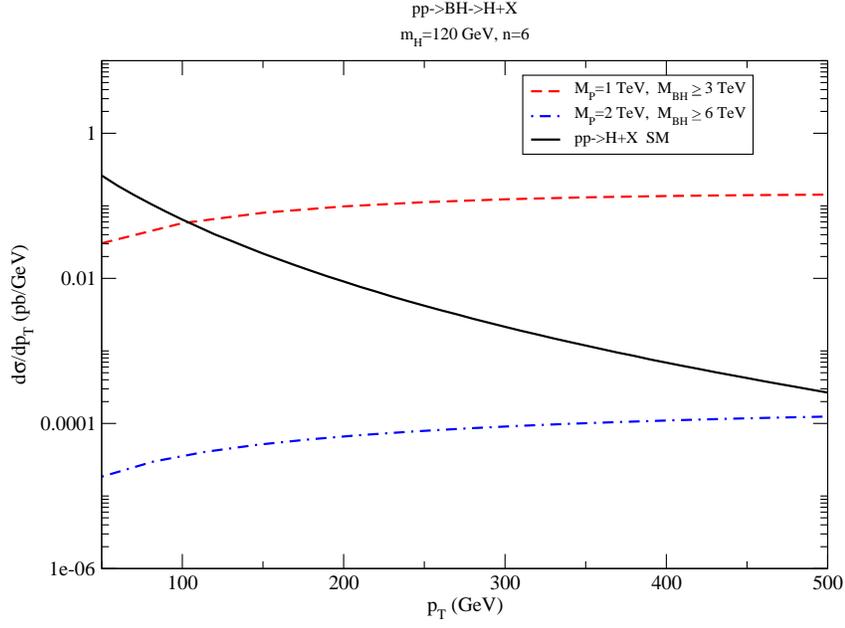,width=3.8in,angle=270}
\end{center}
\caption{Same as Fig.2 but for different values of the fundamental Planck masses and the minimum black hole masses; 
$M_P=1 TeV$ and $M^{min}_{\rm BH}=3 TeV$ (dashed), $M_P=2 TeV$ and $M^{min}_{\rm BH}=6 TeV$ (dot-dashed).}
\label{}
\end{figure}

\begin{figure}[ht]
\begin{center}
\epsfig{file=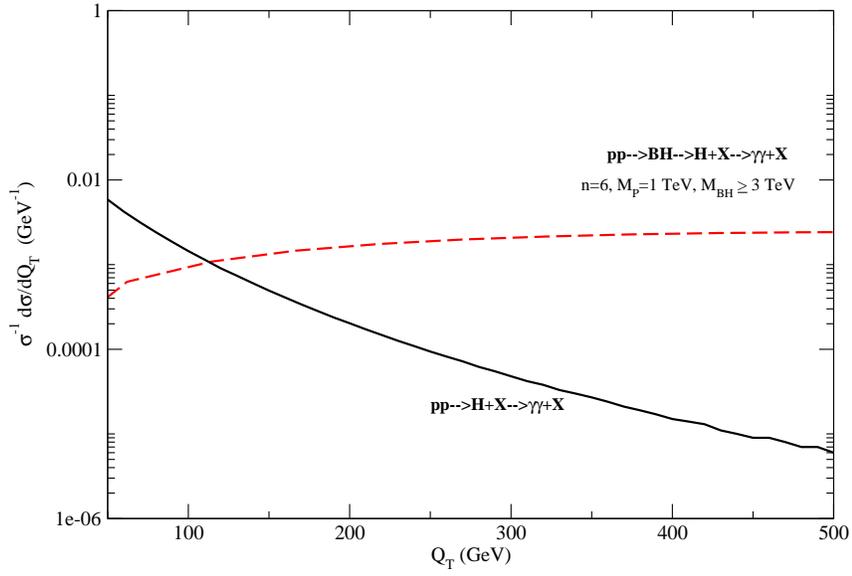,width=3.8in,angle=270}
\end{center}
\caption{Normalized transverse momentum 
 distribution of diphoton pairs, when diphotons are emitted from the Higgs, produced in the 
Standard Model processes (solid line) \cite{smith,smithr} and when 
the Higgs is emitted from black holes (dashed line).} 
\label{}
\end{figure}

\begin{figure}[ht]
\begin{center}
\epsfig{file=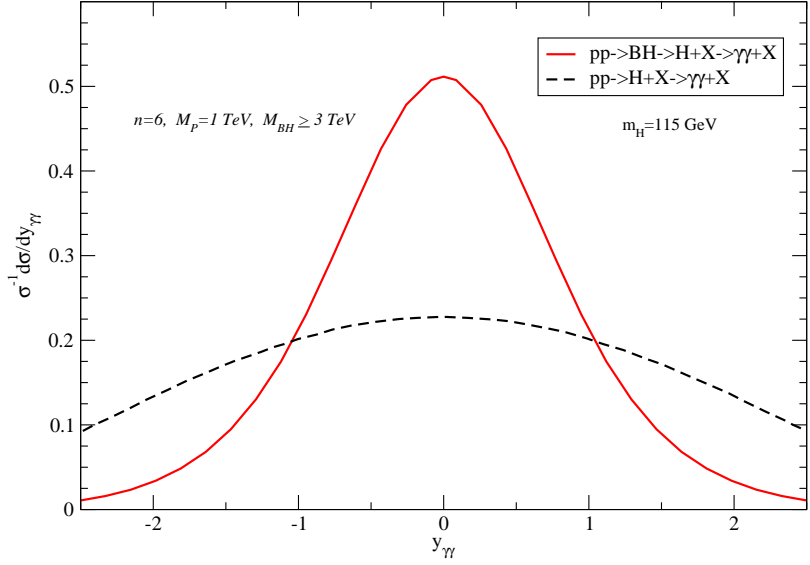,width=3.8in,angle=270}
\end{center}
\caption{Normalized rapidity distribution of the diphoton pairs, 
 when diphotons are emitted from the Higgs, produced in the 
Standard Model processes (solid line) \cite{Dixon} and when 
the Higgs is emitted from black holes (dashed line).} 
\label{}
\end{figure}

\begin{figure}[ht]
\begin{center}
\epsfig{file=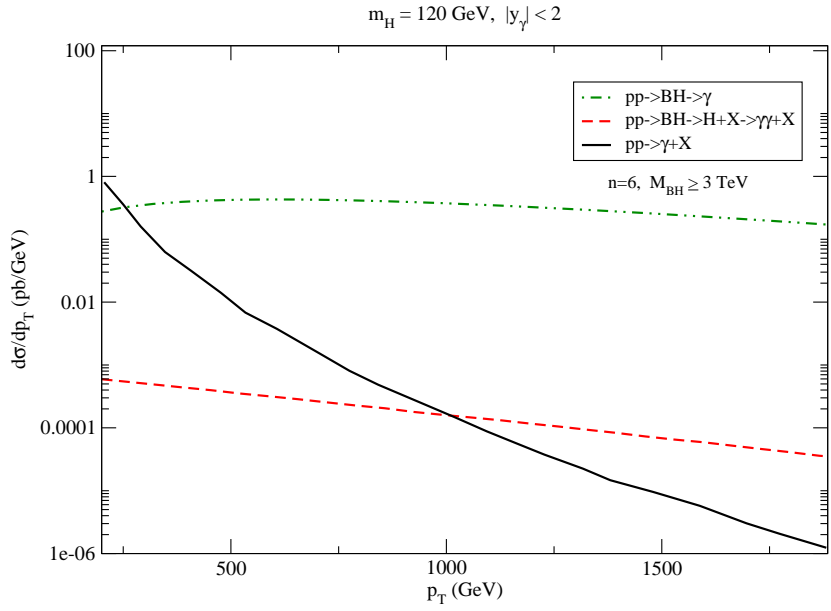,width=3.8in,angle=270}
\end{center}
\caption{Photon transverse momentum distribution, when photons are emitted from black 
holes (dot-dot-dashed line), when they are produced in Higgs decay, when the Higgs is emitted 
from black holes (dashed line) compared to the photons produced in the Standard Model \cite{Blair}.  }
\label{}
\end{figure}

\begin{figure}[ht]
\begin{center}
\epsfig{file=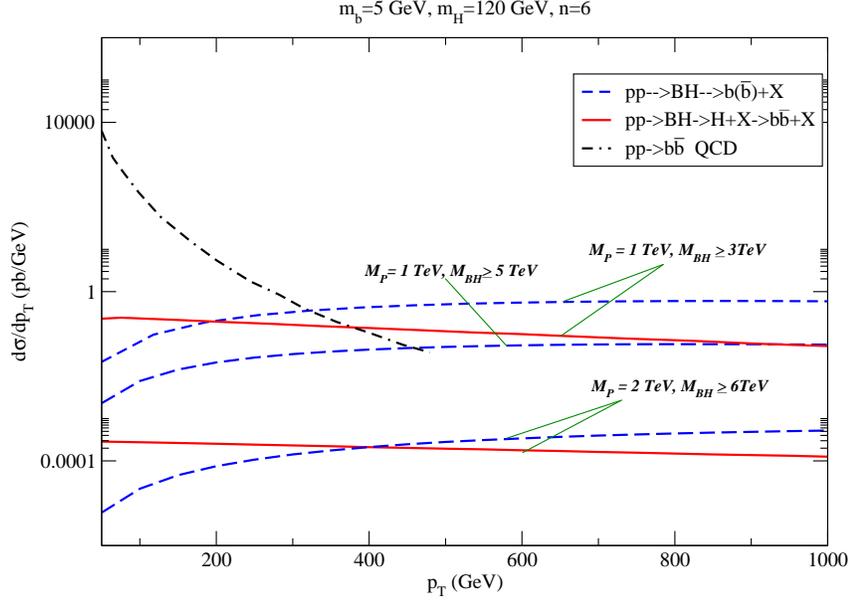,width=3.8in,angle=270}
\end{center}
\caption{Transverse momentum distribution for 
the bottom quarks produced from 
Higgs decay, when the Higgs is emitted from black holes (solid line), 
produced directly from black hole evaporation (dashed line) at the LHC.  
The upper (lower) curves correspond to the Planck mass of 1(2) TeV and 
minimum black hole mass of 3(6) TeV.  The QCD prediction for bottom quark 
production is also presented (dashed-dot-dashed) \cite{Nason}.}
\label{}
\end{figure}

\begin{figure}[]
\begin{center}
\epsfig{file=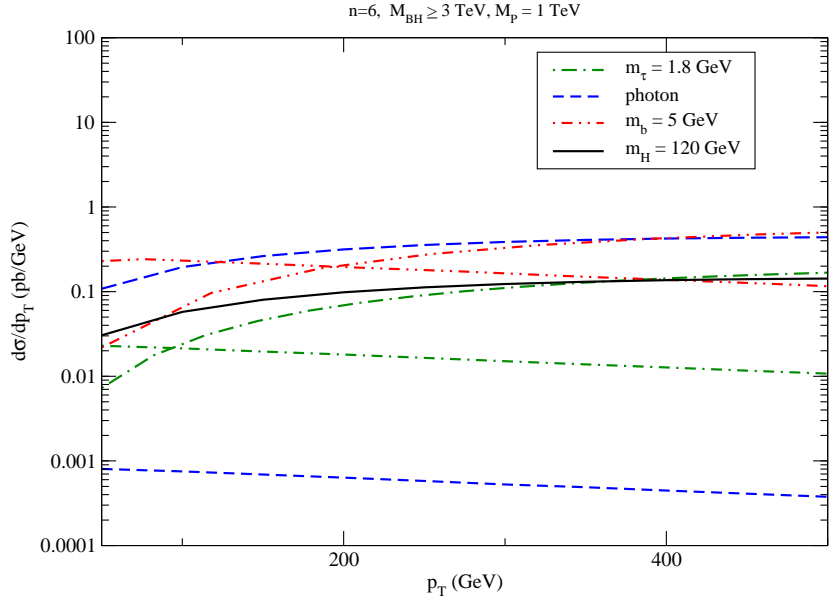,width=3.8in,angle=270}
\end{center}
\caption{Transverse momentum distribution distribution for the bottom 
quarks (dot-dot-dashed line), $\tau$ lepton (dot-dashed line), the Higgs (solid line) 
and single photon (dashed line). The upper curves at large $p_T$ for each particle 
correspond to the direct production from black hole 
evaporation whereas the lower ones corespond to the decay of Higgs from black holes.}
\label{}
\end{figure}

\begin{figure}[ht]
\begin{center}
\epsfig{file=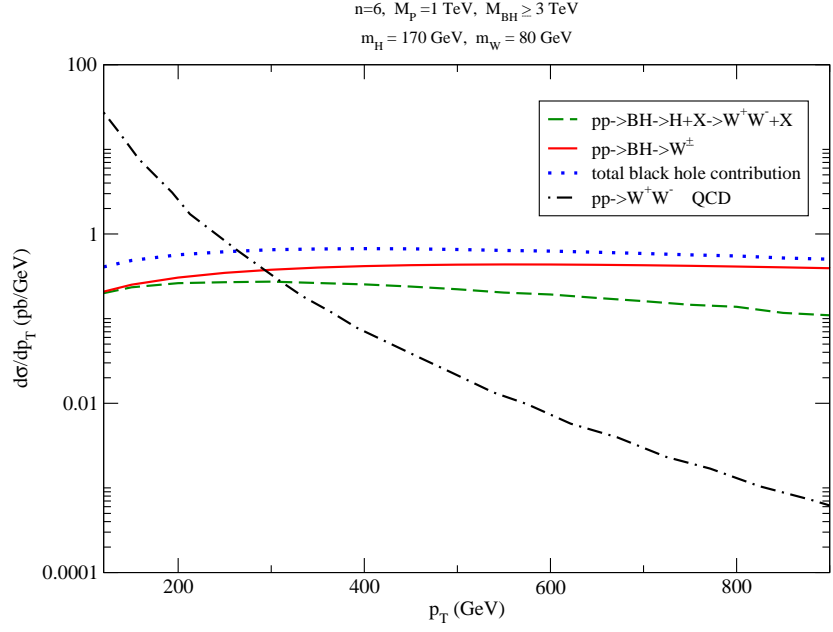,width=3.8in,angle=270}
\end{center}
\caption{Transverse momentum distributions for W boson produced from the  
the black hole (solid line) and from the Higgs, when Higgs is emitted from black holes (dashed line),  
compared with the Standard Model prediction (dash-dotted line) \cite{Wboson}, the total black hole contribution is also shown (dotted line).}  
\label{}
\end{figure}

\begin{figure}[ht]
\begin{center}
\epsfig{file=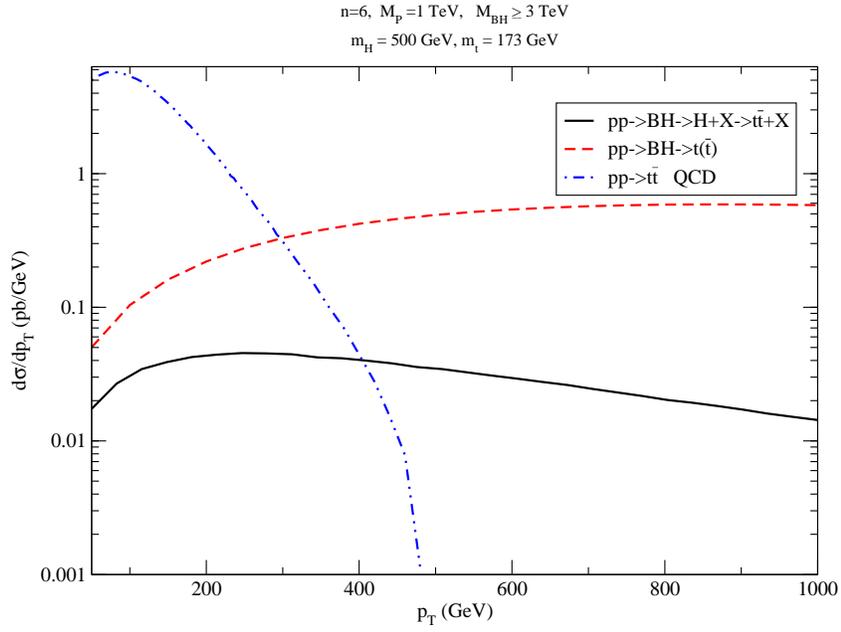,width=3.8in,angle=270}
\end{center}
\caption{Transverse momentum distributions for top quark emitted by black holes (dashed line) and from 
 Higgs decay, when Higgs is emitted from black holes (solid line) compared with the QCD prediction 
(dotted line) \cite{topQCD}.} 
\label{}
\end{figure}

\newpage

\end{document}